\documentstyle{article}
\begin{document}
\setcounter{page}{0}
\null
\vfill
\begin{center}
{\LARGE {\bf Higher Derivatives}} \\
\vskip 1em
{\LARGE {\bf and}} \\
\vskip 1em
{\LARGE {\bf Canonical Formalisms}}
\vfill
{\large \lineskip .75em Takao NAKAMURA  \\ and \\ Shinji HAMAMOTO} \\
\vskip 1.5em
{\it Department of Physics, Toyama University} \\{\it Toyama 930, JAPAN}
\vfill
\null
\abstract
Path integral expressions for three canonical formalisms -- {\it
Ostrogradski's} one, {\it constrained} one and {\it generalized} one -- of
higher-derivative theories are given. For each fomalism we consider both
nonsingular and singular cases. It is shown that three formalisms share the
same path integral expressions. In paticular it is pointed out that the
generalized canonical formalism is connected with the constrained one by a
canonical transformation.
\end{center}
\vfill
\thispagestyle{empty}
\newpage
\section{Introduction}
Higher-derivative theories appear in various scenes of physics,${}^{1),2)}$.
Higher-derivate terms occur as quantum corrections; nonlocal theories, e.g.
string theories, are essentially higher-derivative theories; Einstein gravity
supplemented by curvature squared terms has attracted attention because of its
renormalizability.${}^{3)}$

A canonical formalism for higher-derivative theories was first developed by
Ostrogradski about one and a half centuries ago.${}^{4)}$ He treated only
nonsingular cases, where the Hessian matrices of Lagrangians with respect to
highest derivatives are nonsingular. For singular cases, Dirac's
algorithm${}^{5)}$ for constrained Hamiltonian systems was shown to be
applicable.${}^{6),7)}$ Though being self-consistent, these formulations for
nonsingular and singular cases look different from the conventional canonical
formalism: highest derivatives are discriminated from lower ones, only the
highest ones enjoying Legendre transformations. If we regard the original
higher-derivative systems as equivalent first-derivative systems with
constraints and apply the Dirac's algolithm to the latter ones, we could give
the foundation of the ordinary canonical formalism to the Ostrogradski's
canonical one. This program, constrained canonical formulation of
higher-derivative theories, has bee!
 n carried out in Refs. 6) and 8) for both nonsingular and singular cases. A
generalization of the constrained canonical formalism has been discussed in
Ref. 9).

In all these approaches the sets of canonical equations provided by the
respective formalisms have mainly been considered, and their equivalence to the
set of Euler-Lagrange equations has been shown. To go to quantum theory,
however, the equivalence of the sets of equations of motion is not enough. We
have to confirm the equivalence of off-shell imformation. That is, comparing
path integral expressions of the respective formalisms is essentially
important. This is the subject of the present paper. We give path integral
expressions for each formalism and show they are equivalent to one another. In
paticular it is pointed out that the generalized canonical formalism is
connected with the constrained canonical one by a canonical transformation.

In \S 2, path integral expressions of the Ostrogradski's canonical formalism
are given for both singular and nonsingular cases. In \S 3, path integral
expressions of the constrained canonical formalism are given and it is shown
that the constrained one is equivalent to the Ostrogradski's one. In \S 4, path
integral expressions of the generalized canonical formalism are given. A
further generalization of the formalism described in Ref. 9) is developed. It
is shown by doing a canonical transformation that the generalized one is
equivalent to the Ostrogradski's. Section 5 gives summary and descussion.

\section{Ostrogradski's canonical formalism}

We consider a system described by a generic Lagrangian which contains up to
$n_a$-th derivative of $x_a(t) \, (a=1,\cdots,N)$
\begin{equation}
L=L(x_a,\dot{x}_a,\ddot{x}_a,\cdots ,x_a^{(n_a)}),
\end{equation}
where
\begin{equation}
x_a^{(r_a)} \stackrel{\rm def}{\equiv} \frac{d^{r_a}x_a}{dt^{r_a}}. \qquad
(r_a=1,\cdots,n_a)
\end{equation}
The canonical formalism of Ostrogradski regards $x_a^{(s_a)} (s_a=1,\cdots
,n_a-1)$ as independent coordinates $q_a^{s_a+1}$:
\begin{eqnarray}
x_a^{(s_a)} &\to& q_a^{s_a+1}, \\
L(x_a,\dot{x}_a,\cdots ,x_a^{(n_a)}) &\to& L_{\rm q}(q_a^1,\cdots
,q_a^{n_a},\dot{q}_a^{n_a}).
\end{eqnarray}
The momenta conjugate to $q_a^{n_a}$ is defind as usual by
\begin{equation}
p_a^{n_a} \stackrel{\rm def}{\equiv} \frac{\partial L_{\rm q}}{\partial
\dot{q}_a^{n_a}}. \label{eq:Op}
\end{equation}
The {\it Hessian matrix} of $L_{\rm q}$ is defined by
\begin{equation}
A_{ab} \stackrel{\rm def}{\equiv} \frac{\partial ^2L_{\rm q}}{\partial
\dot{q}_a^{n_a}\partial \dot{q}_b^{n_b}}.
\end{equation}
We say that the system is nonsingular if ${\rm det}A_{ab} \neq 0$, while
singular if ${\rm det}A_{ab}=0 $.

{\bf Nonsingular case }$({\rm det}A_{ab} \neq 0)$

In this case, the relation (\ref{eq:Op}) can be inverted to give
$\dot{q}_a^{n_a}$ as functions of $q^r(r=1,\cdots ,n)$ and $p^n$ :
\begin{equation}
\dot{q}_a^{n_a}=\dot{q}_a^{n_a}(q^r,p^n).
\end{equation}
The Hamiltonian is defined by
\begin{equation}
H_{\rm O} \stackrel{\rm def}{\equiv} p_a^{s_a}q_a^{s_a+1} +
p_a^{n_a}\dot{q}_a^{n_a}(q^r,p^n) - L_{\rm q}\left(q^r,\dot{q}^n(q^r,p^n)
\right). \label{eq:HO}
\end{equation}

It is seen that this construction of the Hamiltonian has several peculiarities
from the view point of the ordinary Legendre transformation:
\begin{enumerate}
\item What appears in Eq.(\ref{eq:HO}) is just a function $L_{\rm q}(q^1,\cdots
,q^n,\dot{q}^n)$ whose Euler derivatives do not produce any meaningful
equations of motion.
\item The momenta $p^s \; (s=1, \cdots ,n-1)$ are multiplied by $q^{s+1}$ not
by $\dot{q}^s$.
\item The momenta $p^s(s=1,\cdots ,n-1)$ are not defined from the Lagrangian
through relations like $\frac{\partial L}{\partial \dot{q}^s}$, but just
introduced as independent canonical variables; only $p^n$'s enjoy special
treatment, defined by Eq.(\ref{eq:Op}) as usual.
\end{enumerate}

Time development of the system is described by the canonical equations of
motion: $\dot{q}=\frac{\partial H_{\rm O}}{\partial p}$, $ \dot{p} =
-\frac{\partial H_{\rm O}}{\partial q}$. That suggests the path integral is
given by
\begin{equation}
Z_{\rm O}=\int {\cal D}q_a^{r_a}{\cal D}p_a^{r_a}\exp i \int dt
[p_a^{r_a}\dot{q}_a^{r_a} - H_{\rm O}(p^r,q^r)].
\end{equation}
At this stage we do not enter into the problem whether or not this expression
can be well-defined. Integrations with respect to $p_a^{s_a}(s_a=1,\cdots
,n_a-1)$ offer a factor $\prod_{s_a=1}^{n_a-1}{\rm \delta}(\dot{q}_a^{s_a} -
q_a^{s_a+1})$ . We can further integrate with respect $q_a^{s_a+1}$, obtaining
\begin{equation}
Z_{\rm O}=\int {\cal D}q_a^1{\cal D}p_a^{n_a} \exp i \int dt
[p_a^{n_a}q_a^{1(n_a)} - \hat{H}_{\rm O}(q^1,q^{1(s)},p^n)], \label{eq:ZO}
\end{equation}
where
\begin{equation}
\hat{H}_{\rm O}(q^1,q^{1(s)},p^n) \stackrel{\rm def}{\equiv}
p_a^{n_a}\dot{q}_a^{n_a}(q^1,q^{1(s)},p^n) - L_{\rm
q}\left(q^1,q^{1(s)},\dot{q}^n(q^1,q^{1(s)},p^n) \right),
\end{equation}
\begin{equation}
q_a^{1(s_a)} \stackrel{\rm def}{\equiv} \frac{d^{s_a}q_a^1}{dt^{s_a}}.
\end{equation}
{\bf Singular case}$(\det A_{ab}=0$, ${\rm rank}A_{ab} = N-\rho )$

In this case, the relation (\ref{eq:Op}) can not be inverted. We have $\rho$
primary constraints:
\begin{equation}
\phi_A(q^r,p^n) \approx 0 , \qquad  (A=1,\cdots ,\rho)   \label{eq:Ophi}
\end{equation}
such that
\begin{equation}
{\rm det}\{\phi_A, \phi_B \}_{\rm P} \neq 0.
\end{equation}
By using Lagrange multipliers $\lambda_A$, we define the Hamiltonian as usual:
\begin{equation}
\bar{H}_{\rm S}(q^r,p^r)=H_{\rm S}(q^r,p^r) + \lambda_A\phi_A(q^r,p^n),
\label{eq:HOS}
\end{equation}
where
\begin{equation}
H_{\rm S}(q^r,p^r) \stackrel{\rm def}{\equiv} p_a^{s_a}q_a^{s_a+1} +
p_a^{n_a}\dot{q}_a^{n_a} - L_{\rm q}(q^r,\dot{q}^n).
\end{equation}
Since $\det \{\phi_A , \phi_B \}_{\rm P} \neq 0$, the primary constraints
(\ref{eq:Ophi}) are second-class ones. The consistency of the primary
constraints (\ref {eq:Ophi}) under their time developments determines all the
Lagrange multipliers $\lambda_A$. The path integral is
\begin{equation}
Z_{\rm Os}=\int {\cal D}q_a^{r_a}{\cal D}p_a^{r_a}{\rm
det}^{\frac{1}{2}}\{\phi_A,\phi_B \}_{\rm P} {\rm \delta}\left(\phi_A(q^r,p^n)
\right)\exp i \int dt [p_a^{r_a}\dot{q}_a^{r_a} - H_{\rm S} ].
\end{equation}
Integrations with respect to $p_a^{s_a}$ and $q_a^{s_a+1}$ give
\begin{equation}
Z_{\rm Os}=\int {\cal D}q_a^1{\cal D}p_a^{n_a}{\rm
det}^{\frac{1}{2}}\{\phi_A,\phi_B \}_{\rm P} {\rm \delta}\left(\phi_A(q^r,p^n)
\right) \exp i \int dt [p_a^{n_a}q_a^{1(n_a)} - \hat{H}_{\rm
S}(q^1,q^{1(s)},p^n)], \label{eq:ZOS}
\end{equation}
where
\begin{equation}
\hat{H}_{\rm S}(q^1,q^{1(s)},p^n) \stackrel{\rm def}{\equiv}
p_a^{n_a}\dot{q}_a^{n_a} - L_{\rm q}(q^1,q^{1(s)},\dot{q}^{n}).
\end{equation}
\section{Constrained canonical formalism}

It has been seen that the Ostrogradski's formalism gives special treatment to
the highest derivatives $q_a^{n_a}$. To treat all the derivatives equally, we
introduce Lagrangian multipriers $\mu_a^{s_a}$ and start with the following
Lagrangian:
\begin{equation}
L_{\rm D}(q^r,\dot{q}^r,\mu^s) \stackrel{\rm def}{\equiv} L_{\rm
q}(q^r,\dot{q}^n) + \mu_a^{s_a}(\dot{q}_a^{s_a} - q_a^{s_a+1}) \label{eq:LD}.
\end{equation}
The conjugate momenta
\begin{eqnarray}
\pi_a^{s_a} &\stackrel{\rm def}{\equiv}& \frac{\partial L_{\rm D}}{\partial
\dot{\mu}_a^{s_a}}=0, \\
p_a^{s_a} &\stackrel{\rm def}{\equiv}& \frac{\partial L_{\rm D}}{\partial
\dot{q}_a^{s_a}}=\mu_a^{s_a},  \\
p_a^{n_a} &\stackrel{\rm def}{\equiv}& \frac{\partial L_{\rm D}}{\partial
\dot{q}_a^{n_a}}=\frac{\partial L_{\rm q}}{\partial \dot{q}_a^{n_a}}
\label{eq:Dpn}
\end{eqnarray}
provide the following primary constraints:
\begin{eqnarray}
&\pi_a^{s_a}& \approx 0 \label{eq:Dpi}, \\
&\psi_a^{s_a} \stackrel{\rm def}{\equiv} p_a^{s_a} - \mu_a^{s_a}& \approx 0.
\label{eq:Dpsi}
\end{eqnarray}

{\bf Nonsingular case}$({\rm det}A_{ab} \neq 0)$

In this case, the relation (\ref{eq:Dpn}) can be inverted to give
$\dot{q}_a^{n_a}$ as functions of $q^r$ and $p^n$:
\begin{equation}
\dot{q}_a^{n_a}=\dot{q}_a^{n_a}(q^r,p^n).
\end{equation}
By introducing Lagrange multipliers $\bar{\lambda}_a^{(1)s_a}$ and
$\bar{\lambda}_a^{(2)s_a}$, the Hamiltonian is defined by
\begin{equation}
\bar{H}_{\rm D}(q^r,p^r)=\pi_a^{s_a}\dot{\mu}_a^{s_a} +
p_a^{r_a}\dot{q}_a^{r_a} - L_{\rm D} + \bar{\lambda}_a^{(1)s_a}\pi_a^{s_a} +
\bar{\lambda}_a^{(2)s_a}\psi_a^{s_a}.
\end{equation}
This can be rewritten as
\begin{equation}
\bar{H}_{\rm D}(q^r,p^r)=H_{\rm D}(q^r,p^r) + \lambda_a^{(1)s_a}\pi_a^{s_a} +
\lambda_a^{(2)s_a}\psi_a^{s_a}, \label{eq:HD}
\end{equation}
where
\begin{eqnarray}
H_{\rm D}(q^r,p^r) &\stackrel{\rm def}{\equiv}& p_a^{s_a}q_a^{s_a+1} +
p_a^{n_a}\dot{q}_a^{n_a} - L_{\rm q}(q^r,\dot{q}^n), \\
\lambda_a^{(1)s_a} &\stackrel{\rm def}{\equiv}& \bar{\lambda}_a^{(1)s_a} +
\dot{\mu}_a^{s_a}, \\
\lambda_a^{(2)s_a} &\stackrel{\rm def}{\equiv}&
\bar{\lambda}_a^{(2)s_a}+\dot{q}_a^{s_a} - q_a^{s_a +1}.
\end{eqnarray}
The Poisson brackets between the primary constraints (\ref{eq:Dpi}) and
(\ref{eq:Dpsi}) are
\begin{eqnarray}
\{\pi_a^{s_a},\psi_b^{s_b} \}_{\rm P} &=& {\rm \delta}_{ab}{\rm
\delta}_{s_as_b}, \\
{\rm otherwise} &=& 0. \nonumber
\end{eqnarray}
Thus, these primary constraints are of the second class. The path integral is
\begin{equation}
Z_{\rm D}=\int {\cal D}q_a^{r_a}{\cal D}p_a^{r_a}{\cal D}\mu_a^{s_a}{\cal
D}\pi_a^{s_a}{\rm \delta}(\pi^s){\rm \delta}(\psi^s)\exp i \int dt
[p_a^{r_a}\dot{q}_a^{r_a} + \pi_a^{s_a}\dot{\mu}_a^{s_a} - H_{\rm D}].
\end{equation}
Integrations with respect to $\pi_a^{s_a}$ and $\mu_a^{s_a}$ give
\begin{equation}
Z_{\rm D}=\int {\cal D}q_a^{r_a}{\cal D}p_a^{r_a} \exp i \int dt
[p_a^{s_a}(\dot{q}_a^{s_a} - q_a^{s_a+1}) + p_a^{n_a}\left(\dot{q}_a^{n_a} -
\dot{q}_a^{n_a}(q^r,p^n) \right) - L_{\rm q}].
\end{equation}
We can further integrate with respect to $p_a^{s_a}$ and $q_a^{s_a+1}$,
obtaining
\begin{equation}
Z_{\rm D}=\int {\cal D}q_a^1{\cal D}p_a^{n_a} \exp i \int dt
[p_a^{n_a}q_a^{1(n_a)} - \hat{H}_{\rm D}(q^1,q^{1(s)},p^n)], \label{eq:ZD}
\end{equation}
where
\begin{equation}
\hat{H}_{\rm D}(q^1,q^{1(s)},p^n)=p_a^{n_a}\dot{q}_a^{n_a}(q^1,q^{1(s)},p^n) -
L_{\rm q}\left(q^1,q^{1(s)},\dot{q}^n(q^1,q^{1(s)},p^n) \right).
\end{equation}
This shows that the path integral $Z_{\rm D}$ is the same as $Z_{\rm O}$ given
by Eq.(\ref{eq:ZO}).

{\bf Singular case } ($\det A_{ab}=0$, ${\rm rank}A_{ab} =N-\rho$)

In this case, the relation (\ref{eq:Dpn}) provides $\rho$ additional
constraints besides (\ref{eq:Dpi}) and (\ref{eq:Dpsi}):
\begin{equation}
\phi_A(q^r,p^n) \approx 0 \; (A=1,\cdots,\rho)  \label{eq:Dphi}
\end{equation}
such that
\begin{equation}
{\rm det}\{\phi_A,\phi_B \}_{\rm P} \neq 0.
\end{equation}
By using Lagrange multipliers $\lambda_A,\lambda_a^{(1)s_a}$ and
$\lambda_a^{(2)s_a}$, the Hamiltonian is defined by
\begin{equation}
\bar{H}_{\rm Ds}(q^r,p^r)=H_{\rm D}(q^r,p^r) + \lambda_a^{(1)s_a}\pi_a^{s_a} +
\lambda_a^{(2)s_a}\psi_a^{s_a} + \lambda_A\phi_A,
\end{equation}
where
\begin{equation}
H_{\rm Ds}(q^r,p^r) \stackrel{\rm def}{\equiv} p_a^{s_a}q_a^{s_a+1} +
p_a^{n_a}\dot{q}_a^{n_a} - L_{\rm q}(q^r,\dot{q}^n).
\end{equation}

The Poisson brackets between the primary constraints are
\begin{eqnarray}
\{\pi_a^{s_a},\psi_b^{s_b} \}_{\rm P} &=& {\rm \delta}_{ab}{\rm
\delta}_{s_as_b}, \\
\{\psi_a^{s_a},\phi_B \}_{\rm P} &=& -\frac{\partial \phi_B}{\partial
q_a^{s_a}}, \\
\{\phi_A,\phi_B \}_{\rm P} &\stackrel{\rm def}{\equiv}& c_{AB}, \\
{\rm otherwise} &=& 0. \nonumber
\end{eqnarray}
All the constraints $\Phi_\alpha \stackrel{\rm def}{\equiv}
(\pi_a^{s_a},\psi_a^{s_a},\phi_A)$ form a set of second-class constraints
because the determinant of the matrix $\left(\{\Phi_\alpha,\Phi_\beta \}_{\rm
P} \right)$ is non-zero:
\begin{equation}
{\rm det}\{\Phi_\alpha,\Phi_\beta \}_{\rm P}={\rm det}c_{AB} \neq 0.
\end{equation}
The consistency of these constraints under their time developments fixes all
the Lagrange multipliers.
The path integral is
\begin{equation}
Z_{\rm Ds}=\int {\cal D}q_a^{r_a}{\cal D}p_a^{r_a}{\cal D}\mu_a^{s_a}{\cal
D}\pi_a^{s_a}{\rm det}^{\frac{1}{2}}c_{AB}{\rm \delta}(\pi_a^{s_a}){\rm
\delta}(\psi_a^{s_a}){\rm \delta}(\phi_A)\exp i \int dt
[p_a^{r_a}\dot{q}_a^{r_a} + \pi_a^{s_a}\dot{\mu}_a^{s_a} - H_{\rm Ds}].
\end{equation}
Integrations with respect to $\mu_a^{s_a},\pi_a^{s_a},p_a^{s_a}$ and
$q_a^{s_a+1}$ give
\begin{equation}
Z_{\rm Ds}=\int {\cal D}q_a^1{\cal D}p_a^{n_a}{\rm det}^{\frac{1}{2}}c_{AB}
{\rm \delta}(\phi_A)\exp i \int dt [p_a^{n_a}q_a^{1(n_a)} - \hat{H}_{\rm
Ds}(q^1,q^{1(s)},p^n)], \label{eq:ZDS}
\end{equation}
where
\begin{equation}
\hat{H}_{\rm Ds}(q^1,q^{1(s)},p^n) \stackrel{\rm def}{\equiv}
p_a^{n_a}\dot{q}_a^{n_a} - L_{\rm q}\left(q^1,q^{1(s)},\dot{q}^n \right).
\end{equation}
This shows that the path integral $Z_{\rm Ds}$ is the same as $Z_{\rm Os}$
given by (\ref{eq:ZOS}).
\section{Generalized canonical formalism}

In this section we consider a further generalization of the formalism described
in ~Ref. 9).

We regard $x_a^{(s_a)}$ and $x_a^{(n_a)}$ as independent coordinates
$q_a^{s_a+1}$ and $v_a$ respectively:
\begin{eqnarray}
x_a^{(s_a)} &\to& q_a^{s_a+1}, \\
x_a^{(n_a)} &\to& v_a, \\
L(x,\dot{x},\ddot{x},\cdots,x^{(n)}) &\to& L_{\rm q}(q^1,\cdots,q^n,v).
\end{eqnarray}
The other generalized coordinates $Q_a^{r_a}$ are introduced as arbitary
functions of $q^r$
\begin{equation}
Q_a^{r_a}=Q_a^{r_a}(q^r) \label{eq:Qr}
\end{equation}
shch that
\begin{equation}
\det \frac{\partial Q_b^{r_b}}{\partial q_a^{r_a}} \neq 0.
\end{equation}
Eq. (\ref{eq:Qr}) can be inverted to give $q^r$ as functions of $Q^r$:
\begin{equation}
q_a^{r_a}=q_a^{r_a}(Q^r). \label{eq:Sqr}
\end{equation}
Defferentiating Eq. (\ref{eq:Qr}) and (\ref{eq:Sqr}) with respect to time gives
\begin{eqnarray}
\dot{q}_a^{r_a} &=& \dot{Q}_b^{r_b}\frac{\partial q_a^{r_a}(Q^r)}{\partial
Q_b^{r_b}}, \\
\dot{Q}_a^{r_a} &=& \dot{q}_b^{r_b}\frac{\partial Q_a^{r_a}(q^r)}{\partial
q_b^{r_b}}.
\end{eqnarray}
We introduce new variables defined by
\begin{equation}
V_a \stackrel{\rm def}{\equiv} q_b^{s_b+1}\frac{\partial Q_a^{n_a}}{\partial
q_b^{s_b}} + v_b\frac{\partial Q_a^{n_a}}{\partial q_b^{n_b}}, \label{eq:V}
\end{equation}
where we assume that $Q_a^{n_a}$'s satisfy
\begin{equation}
\det\frac{\partial Q_b^{n_b}}{\partial q_a^{n_a}} \neq 0.
\end{equation}
Eq. (\ref{eq:V}) can be inverted with respect to $v$ as
\begin{equation}
v_a=\left(\frac{\partial Q_b^{n_b}}{\partial q_a^{n_a}} \right)^{-1}\left(V_b -
q_c^{s_c+1}\frac{\partial Q_b^{n_b}}{\partial q_c^{s_c}} \right).
\end{equation}
Functions $\bar{Q}_a^{s_a}$ are defined by
\begin{equation}
\bar{Q}_a^{s_a} \stackrel{\rm def}{\equiv} \left(q_b^{s_b+1}\frac{\partial
Q_a^{s_a}}{\partial q_b^{s_b}} + v_b\frac{\partial Q_a^{s_a}}{\partial
q_b^{n_b}} \right)|_{v=v(Q,V)}^{q=q(Q)}. \label{eq:Qbar}
\end{equation}
We introduce Lagrange multipliers $M_a^{r_a}$ and start from the following {\it
generalized Lagrangian}:
\begin{equation}
L_{\rm G}(Q^r,\dot{Q^r},V,M^r) \stackrel{\rm def}{\equiv} L_{\rm Q}(Q^r,V) +
M_a^{s_a}(\dot{Q}_a ^{s_a} - \bar{Q}_a^{s_a}) + M_a^{n_a}(\dot{Q}_a^{n_a} -
V_a), \label{eq:LG}
\end{equation}
where
\begin{equation}
L_{\rm Q}(Q^r,V) \stackrel{\rm def}{\equiv} L_{\rm
q}(q^r,v)|_{v=v(Q,V)}^{q=q(Q)}.
\end{equation}

Here it is interesting to consider a special case of the generalized
Lagrangian. Choose
\begin{equation}
Q^r=q^r , \;  V = v.
\end{equation}
Then the Lagrangian (\ref{eq:LG}) reduces to
\begin{equation}
L_{\rm g}(q^r,\dot{q}^r,v,\mu^r)=L_{\rm q}(q^r,v) + \mu_a^{s_a}(\dot{q}_a^{s_a}
- q_a^{s_a+1}) + \mu_a^{n_a}(\dot{q}_a^{n_a} - v_a). \label{eq:LSg}
\end{equation}
This Lagrangian is similar to the Lagrangian (\ref{eq:LD}), except for term
containing the variables $v$. The equivalence between the two Lagrangians is
proved later.

For the Lagrangian (\ref{eq:LG}) the conjugate momenta
\begin{eqnarray}
\Pi_a^{r_a} &\stackrel{\rm def}{\equiv}& \frac{\partial L_{\rm G}}{\partial
\dot{M}_a^{r_a}}=0, \\
P_a^{r_a} &\stackrel{\rm def}{\equiv}& \frac{\partial L_{\rm G}}{\partial
\dot{Q}_a^{r_a}}=M_a^{r_a}, \\
\Theta_a &\stackrel{\rm def}{\equiv}& \frac{\partial L_{\rm G}}{\partial
\dot{V}_a}=0
\end{eqnarray}
provide the following primary constraints:
\begin{eqnarray}
\Pi_a^{r_a} &\approx& 0,  \label{eq:Gpi}\\
\Psi_a^{r_a} \stackrel{\rm def}{\equiv} P_a^{r_a} - M_a^{r_a} &\approx& 0
\label{eq:Gpsi}, \\
\Theta_a &\approx& 0. \label{eq:GT}
\end{eqnarray}
The consistency of the primary constraints under their time developments
produces a secondary constraint:
\begin{equation}
\Gamma_a \stackrel{\rm def}{\equiv} -P_b^{s_b}\frac{\partial
\bar{Q}_b^{s_b}}{\partial V_a} -P_a^{n_a} + \frac{\partial L_{\rm Q}}{\partial
V_a}. \label{eq:GG}
\end{equation}
By introducing Lagrange multipliers
$\Lambda_a^{(1)r_a},\Lambda_a^{(2)r_a},\Lambda_a^{(3)}$ and $\Lambda_a^{(4)}$,
the Hamilonian is given by
\begin{equation}
\bar{H}_{\rm G}=H_{\rm G}(Q^r,P^r,V) + \Lambda_a^{(1)r_a}\Pi_a^{r_a} +
\Lambda_a^{(2)r_a}\Psi_a^{r_a} + \Lambda_a^{(3)}\Theta_a +
\Lambda_a^{(4)}\Gamma_a,  \label{eq:HGa}
\end{equation}
where
\begin{equation}
H_{\rm G}(Q^r,P^r,V) \stackrel{\rm def}{\equiv} P_a^{s_a}\bar{Q}_a^{s_a} +
P_a^{n_a}V_a - L_{\rm Q}(Q^r,V).
\end{equation}
The Poisson brackets between the constraints are
\begin{eqnarray}
\{\Pi_a^{r_a},\Psi_b^{r_b} \}_{\rm P} &=& \delta_{ab} \delta_{r_ar_b}, \\
\{\Psi_a^{r_a},\Gamma_b \}_{\rm P} &=& P_c^{s_c}\frac{\partial^2
\bar{Q}_c^{s_c}}{\partial Q_a^{r_a}\partial V_b} - \frac{\partial^2 L_{\rm
Q}}{\partial Q_a^{r_a}\partial V_b}, \\
\{\Theta_a,\Gamma_b \}_{\rm P} &=& - \frac{\partial^2 L_{\rm Q}}{\partial
V_a\partial V_b}, \\
\{\Gamma_a,\Gamma_b \}_{\rm P} &\stackrel{\rm def}{\equiv}& C_{ab}, \\
{\rm otherwise} &=& 0. \nonumber
\end{eqnarray}
All the consraints $ \Sigma_\alpha \stackrel{\rm def}{\equiv}
(\Theta_a,\Psi_a^{r_a},\Pi_a^{r_a},\Gamma_a)$ give for the determinant of the
matrix $\left(\{\Sigma_\alpha,\Sigma_\beta \}_{\rm P} \right)$
\begin{equation}
\det \{\Sigma_\alpha,\Sigma_\beta \}_{\rm P}=- {\rm det}^2\frac{\partial^2
L_{\rm Q}}{\partial V_a\partial V_b}.
\end{equation}
Therefore we find that if
\begin{equation}
\det\frac{\partial^2 L_{\rm Q}}{\partial V_a\partial V_b} \neq 0,
\end{equation}
then the system is nonsingular; on the other hand if
\begin{equation}
\det\frac{\partial^2 L_{\rm Q}}{\partial V_a\partial V_b}=0,
\end{equation}
then it is singular.

{\bf Nonsingular case}

In this case, the constraints (\ref{eq:Gpi}) $\sim$ (\ref{eq:GG}) are
second-class ones. Thus the consistency of the constraints under their time
developments fixes all the Lagrange multiplires. The path integral is
\begin{eqnarray}
Z_{\rm G}=\int {\cal D}Q_a^{r_a}{\cal D}P_a^{r_a}{\cal D}M_a^{r_a}{\cal
D}\Pi_a^{r_a}{\cal D}V_a{\cal D}\Theta_a{\rm \delta}(\Pi^r){\rm
\delta}(\Psi^r){\rm \delta}(\Theta){\rm \delta}(\Gamma) \det\frac{\partial^2
L_{\rm Q}}{\partial V_a\partial V_b}  \nonumber \\
\times\exp i  \int dt [P_a^{r_a}\dot{Q}_a^{r_a} + \Pi_a^{r_a}\dot{M}_a^{r_a} +
\Theta_a\dot{V}_a - H_{\rm G}].
\end{eqnarray}
Integrations with respect to $\Pi^r,\Theta,M^r$ give
\begin{eqnarray}
Z_{\rm G} &=& \int {\cal D}Q_a^{r_a}{\cal D}P_a^{r_a}{\cal D}V_a {\rm
\delta}\left(\Gamma (Q^r,P^r,V) \right) \det\frac{\partial^2 L_{\rm
Q}}{\partial V_a\partial V_b} \nonumber \\
& &{}\times\exp i \int dt [P_a^{s_a}(\dot{Q}_a^{s_a} - \bar{Q}_a^{s_a}) +
P_a^{n_a}(\dot{Q}_a^{n_a} - V_a) + L_{\rm Q}].  \label{eq:ZG}
\end{eqnarray}

{\bf Singular case }

In this case, we have extra constraints in addition to (\ref{eq:Gpi}) $\sim$
(\ref{eq:GG}):
\begin{equation}
\Omega_A(Q^r,P^s,V) \approx 0. \label{eq:GSO}
\end{equation}
Then by introducing Lagrange multipliers
$\Lambda_a^{(1)r_a},\Lambda_a^{(2)r_a},\Lambda_a^{(3)},\Lambda_a^{(4)}$ and
$\Lambda_A^{(5)}$, the Hamiltonian is given by
\begin{eqnarray}
\bar{H}_{\rm Gs} &=& H_{\rm G}(Q^r,P^r,V) + \Lambda_a^{(1)r_a}\Pi_a^{r_a}
+\Lambda_a^{(2)r_a}\Psi_a^{r_a} \nonumber \\
& &{}+ \Lambda_a^{(3)}\Theta_a + \Lambda_a^{(4)}\Gamma_a +
\Lambda_A^{(5)}\Omega_A. \label{eq:HGSa}
\end{eqnarray}
The Poisson brackets between the constrains are
\begin{eqnarray}
\{\Pi_a^{r_a},\Psi_b^{r_b} \}_{\rm P} &=& \delta_{ab}\delta_{r_ar_b}, \\
\{\Psi_a^{r_a},\Gamma_b \}_{\rm P} &=& P_c^{s_c}\frac{\partial^2
\bar{Q}_c^{s_c}}{\partial Q_a^{r_a}\partial V_b} - \frac{\partial^2L_{\rm
Q}}{\partial Q_a^{r_a}\partial V_b}, \\
\{\Theta_a,\Gamma_b \}_{\rm P} &=& -\frac{\partial^2 L_{\rm Q}}{\partial
V_a\partial V_b}, \\
\{\Gamma_a,\Gamma_b \}_{\rm P} &\stackrel{\rm def}{\equiv}& C_{ab}, \\
\{\Psi_a^{r_a},\Omega_A \}_{\rm P} &=& -\frac{\partial\Omega_A}{\partial
Q_a^{r_a}}, \\
\{\Theta_a,\Omega_A \}_{\rm P} &=& -\frac{\partial\Omega_A}{\partial V_a}, \\
\{\Gamma_a,\Omega_A \}_{\rm P} &=& \left(-P_b^{s_b}\frac{\partial^2
Q_b^{s_b}}{\partial Q_c^{s_c}\partial V_a} + \frac{\partial^2L_{\rm
Q}}{\partial Q_c^{s_c}\partial V_a} \right)\frac{\partial\Omega_A}{\partial
P_c^{s_c}} + \frac{\partial\bar{Q}_c^{s_c}}{\partial
V_a}\frac{\partial\Omega_A}{\partial Q_c^{s_c}} + \frac{\partial
\Omega_A}{\partial Q_a^{n_a}}, \\
\{\Omega_A,\Omega_B \}_{\rm P} &\stackrel{\rm def}{\equiv}& D_{AB}, \\
{\rm otherwise} &=& 0. \nonumber
\end{eqnarray}
For all the constraints $ \Sigma_\alpha^{\rm (s)} \stackrel{\rm def}{\equiv}
(\Theta_a,\Psi_a^{r_a},\Pi_a^{r_a},\Gamma_a,\Omega_A)$, the determinant of the
matrix $\left(\{\Sigma_\alpha^{\rm (s)},\Sigma_\beta^{\rm (s)} \}_{\rm P}
\right)$ is
\begin{equation}
{\rm det}\{\Sigma_\alpha^{\rm (s)}, \Sigma_\beta^{\rm (s)} \}_{\rm P}=\det
\pmatrix{
\displaystyle 0 & \displaystyle -\frac{\partial^2L_{\rm Q}}{\partial
V_a\partial V_b} & \displaystyle -\frac{\partial\Omega_B}{\partial V_a} \cr
\displaystyle \frac{\partial^2L_{\rm Q}}{\partial V_a\partial V_b} &
\displaystyle C_{ab} & \displaystyle\{\Gamma_a,\Omega_B \}_{\rm P} \cr
\displaystyle\frac{\partial\Omega_A}{\partial V_b} &
\displaystyle\{\Omega_A,\Gamma_B \}_{\rm P} & \displaystyle D_{AB} \cr }.
\end{equation}
If this determinant is nonzero, we assume this is the case, then all the
constraints are of the second class and all the Lagrange multipliers are fixed.
The path integral is
\begin{eqnarray}
Z_{\rm Gs} &=& \int {\cal D}Q^r{\cal D}P^r{\cal D}M^r{\cal D}\Pi^r{\cal
D}V{\cal D}\Theta{\rm \delta}(\Pi^r){\rm \delta}(\Psi^r){\rm
\delta}(\Gamma){\rm \delta}(\Theta){\rm det}^{\frac{1}{2}}\{\Sigma_\alpha^{\rm
(s)},\Sigma_\beta^{\rm (s)} \}_{\rm P} \nonumber \\
& &{} \times \exp i \int dt [P_a^{r_a}\dot{Q}_a^{r_a} +
\Pi_a^{r_a}\dot{M}_a^{r_a} + \Theta_a\dot{V}_a - H_{\rm G} ].
\end{eqnarray}
Integrations with respect to $M_a^{r_a},\Pi_a^{r_a}$ and $\Theta_a$ give
\begin{eqnarray}
Z_{\rm Gs} &=& \int{\cal D}Q^r{\cal D}P^r{\cal D}V{\rm \delta}(\Gamma_a){\rm
\delta}(\Omega_A){\rm det}^{\frac{1}{2}}\{\Sigma_\alpha^{\rm
(s)},\Sigma_\beta^{\rm (s)} \}_{\rm P} \nonumber \\
& &{} \times \exp i\int dt [P_a^{s_a}(\dot{Q}_a^{s_a} - \bar{Q}_a^{s_a}) +
P_a^{n_a}(\dot{Q}_a^{n_a} - V_a) + L_{\rm Q}(Q,V) ]. \label{eq:ZGS}
\end{eqnarray}

Next, we consider the relations between the path integral expressions $Z_{\rm
D}$ (\ref{eq:ZD}) and $Z_{\rm G}$ (\ref{eq:ZG}) (or $Z_{\rm Ds}$ (\ref{eq:ZDS})
and $Z_{\rm G}$ (\ref{eq:ZG})). In fact, these are shown to be connected with
each other through a canonical transformation.

Consider a canonical transformation $(q,p) \to (Q,P)$. The generating function
has the form
\begin{equation}
F(Q,p)=p_a^{r_a}q_a^{r_a}(Q^r), \label{eq:F}
\end{equation}
and gives
\begin{eqnarray}
q_a^{r_a} &=& \frac{\partial F}{\partial p_a^{r_a}}
=q_a^{r_a}(Q^r),\label{eq:qQ} \\
P_a^{r_a} &=& \frac{\partial F}{\partial Q_a^{r_a}} =p_b^{r_b}\frac{\partial
q_b^{r_b}(Q^r)}{\partial Q_a^{r_a}}. \label{eq:pP}
\end{eqnarray}
Eqs. (\ref{eq:qQ}) and (\ref{eq:pP}) can be inverted to give
\begin{eqnarray}
Q_a^{r_a} &=& Q_a^{r_a}(q^r),  \label{eq:QL} \\
p_a^{r_a} &=& P_b^{r_b} \frac{\partial Q_b^{r_b}(q^r)}{\partial q_a^{r_a}}.
\label{eq:PL}
\end{eqnarray}

{\bf Nonsingular case}

We start with the Lagrangian $L_{\rm g}$ (\ref{eq:LSg}). The conjugate momenta
\begin{eqnarray}
\pi_a^{r_a} &\stackrel{\rm def}{\equiv}& \frac{\partial L_{\rm g}}{\partial
\dot{\mu}_a^{r_a}}=0, \\
p_a^{r_a} &\stackrel{\rm def}{\equiv}& \frac{\partial L_{\rm g}}{\partial
\dot{q}_a^{r_a}}=\mu_a^{r_a}, \\
\theta_a &\stackrel{\rm def}{\equiv}& \frac{\partial L_{\rm g}}{\partial
\dot{v}_a}=0
\end{eqnarray}
provide the following primary constraints:
\begin{eqnarray}
&\pi_a^{r_a}& \approx 0,  \label{eq:sgpi}  \\
&\psi_a^{r_a} \stackrel{\rm def}{\equiv} p_a^{r_a} - \mu_a^{r_a}& \approx 0,
\label{eq:sgpsi} \\
&\theta_a& \approx 0. \label{eq:sgt}
\end{eqnarray}
We get the following secondary constraints:
\begin{equation}
\gamma_a \stackrel{\rm def}{\equiv} p_a^{n_a} - \frac{\partial L_{\rm
g}}{\partial v_a}. \label{eq:sgg}
\end{equation}
By introducing Lagrange multipliers
$\lambda_a^{(1)r_a},\lambda_a^{(2)r_a},\lambda_a^{(3)}$ and $\lambda_a^{(4)}$,
the Hamiltonian is given by
\begin{equation}
\bar{H}_{\rm g}=H_{\rm g}(q^r,p^r) + \lambda_a^{(1)r_a}\pi_a^{r_a} +
\lambda_a^{(2)r_a}\psi_a^{r_a} + \lambda_a^{(3)}\theta_a +
\lambda_a^{(4)}\gamma_a,
\end{equation}
where
\begin{equation}
H_{\rm g}(q^r,p^r) \stackrel{\rm def}{\equiv} p_a^{s_a}q_a^{s_a+1} +
p_a^{n_a}v_a - L_{\rm q}(q,v).
\end{equation}
For all the constraints are $\sigma_\alpha \stackrel{\rm def}{\equiv}
(\theta_a,\psi_a^{r_a},\pi_a^{r_a},\gamma_a)$, the determinant of the matrix
$\left(\{\sigma_\alpha,\sigma_\beta \}_{\rm P} \right)$ is
\begin{equation}
{\rm det}\{\sigma_\alpha,\sigma_\beta \}_{\rm P}=-{\rm det}^2\frac{\partial^2
L_{\rm q}}{\partial v_a\partial v_b}.
\end{equation}
If this determinant is nonzero, then all the Lagrange multipliers are
determined.  The path integral is
\begin{eqnarray}
Z_{\rm g} &=& \int {\cal D}q_a^{r_a}{\cal D}p_a^{r_a}{\cal D}\mu_a^{r_a}{\cal
D}\pi_a^{r_a}{\cal D}v_a{\cal D}\theta_a{\rm \delta}(\pi^r){\rm
\delta}(\psi^r){\rm \delta}(\theta){\rm \delta}(\gamma){\rm
det}\frac{\partial^2L_{\rm q}}{\partial v_a \partial v_b}  \nonumber \\
& & {} \times \exp i \int dt [p_a^{r_a}\dot{p}_a^{r_a} +
\pi_a^{r_a}\dot{\mu}_a^{r_a} + \theta_a\dot{v}_a - \bar{H}_{\rm g}].
\end{eqnarray}
Integrations with respect to $\mu^r,\pi^r$ and $\theta$ give
\begin{eqnarray}
Z_{\rm g} &=& \int {\cal D}q_a^{r_a}{\cal D}p_a^{r_a}{\cal D}v_a{\rm
\delta}(\gamma_a){\rm det}\frac{\partial^2 L_{\rm q}}{\partial v_a \partial
v_b} \nonumber \\
& &{} \times \exp i \int dt [p_a^{s_a}(\dot{q}_a^{s_a} - q_a^{s_a+1}) +
p_a^{n_a}(\dot{q}_a^{n_a} - v_a) + L_{\rm q}].\label{eq:120}
\end{eqnarray}
We can futher integrate with respect $p_a^{s_a},q_a^{s_a+1}$ and $v_a$,
obtaining
\begin{equation}
Z_{\rm g}=\int {\cal D}q_a^1{\cal D}p_a^{n_a} \exp i \int dt
[p_a^{n_a}q_a^{1(n_a)} - \hat{H}_{\rm g}(q^1,q^{1(s)},p^n)], \label{eq:sZg}
\end{equation}
where
\begin{equation}
\hat{H}_{\rm g}(q^1,q^{1(s)},p^n) \stackrel{\rm def}{\equiv}
p_a^{n_a}v_a(q^1,q^{1(s)},p^n) - L_{\rm
q}\left(q^1,q^{1(s)},v(q^1,q^{1(s)},p^n) \right).
\end{equation}
Putting $v_a=\dot{q}_a^{n_a}$ in this equation shows that the path integral
$Z_{\rm g}$ is the same as $Z_{\rm O}$ given by (\ref{eq:ZO}) (and also $Z_{\rm
D}$ in (\ref{eq:ZD})).

Next, by doing the canonical transformation generated by $F$ in (\ref{eq:F}),
we show that the path integral $Z_{\rm g}$ is equivalent to $Z_{\rm G}$ given
by (\ref{eq:ZG}). Referring to Eqs. (\ref{eq:qQ}) $\sim$ (\ref{eq:PL}) and
(58), the following relation is inserted into $Z_{\rm g}$ in Eq.
(\ref{eq:120}):
\begin{eqnarray}
\int {\cal D}Q_a^{r_a}{\cal D}P_a^{r_a}{\cal D} V_a {\rm \delta}
\left(q_a^{r_a} - q_a^{r_a}(Q^{r}) \right){\rm \delta}\left(p_a^{r_a} -
P_b^{r_b}\frac{\partial Q_b^{r_b}}{\partial q_a^{r_a}} \right)
\qquad
\qquad
\qquad
\nonumber \\%
\qquad
\qquad
\qquad
\times
{\rm det}\left(\frac{\partial Q_a^{n_a}}{\partial q_b^{n_b}} \right)^{-1} {\rm
\delta}\left(v_b - \left(\frac{\partial Q_a^{n_a}}{\partial q_b^{n_b}}
\right)^{-1}\left(V_a - q_c^{s_c+1}\frac{\partial Q_a^{n_a}}{\partial
q_c^{s_c}} \right) \right)=1. \label{eq:unity}
\end{eqnarray}
Then we have
\begin{eqnarray}
Z_{\rm g}=\int {\cal D}q^r{\cal D}p^r{\cal D}v{\cal D}Q^r{\cal D}P^r{\cal
D}V{\rm \delta}\left(q^r - q^r(Q^r) \right){\rm \delta}\left(p^r - p^r(Q^r,P^r)
\right){\rm \delta}\left(v - v(Q^r,V) \right){\rm \delta}(\gamma) \nonumber \\
\times{\rm det}\left(\frac{\partial Q_a^{n_a}}{\partial q_b^{n_b}}
\right)^{-1}{\rm \det}\frac{\partial^2L_{\rm q}}{\partial v_a \partial v_b}
\exp i \int dt [p_a^{s_a}(\dot{q}_a^{s_a} - q_a^{s_a+1}) +
p_a^{n_a}(\dot{q}_a^{n_a}-v_a) + L_{\rm q} ].
\end{eqnarray}
Integrations with respect to $q^r,p^r$ and $v$ give
\begin{eqnarray}
Z_{\rm g}=\int {\cal D}Q_a^{r_a}{\cal D}P_a^{r_a}{\cal D}V_a\left[{\rm
\delta}\left(\frac{\partial L_{\rm q}}{\partial v_a} -P_b^{r_b}\frac{\partial
Q_b^{r_b}}{\partial q_a^{n_a}} \right) {\rm det}\frac{\partial^2 L_{\rm
q}}{\partial v_a \partial v_b }{\rm det}\left(\frac{\partial
Q_a^{n_a}}{\partial q_b^{r_b}} \right)^{-1} \right]|_{v=v(Q,V)}^{q^r=q^r(Q)}
\nonumber \\
\times \exp i \int dt [P_b^{r_b}\frac{\partial Q_b^{r_b}}{\partial
q_a^{r_a}}\dot{Q}_c^{r_c}\frac{\partial q_a^{r_a}}{\partial Q_c^{r_c}} -
P_c^{r_c}\frac{\partial Q_c^{r_c}}{\partial
q_b^{s_b}}q_b^{s_b+1}(Q)-P_c^{r_c}\frac{\partial Q_c^{r_c}}{\partial
q_b^{n_b}}v_b(Q^r,V) + L_{\rm Q}].
\end{eqnarray}
By using (\ref{eq:V}),(\ref{eq:Qbar}) and the relations
\begin{eqnarray}
{\rm \delta}\left(\gamma_a(q^r,p^n,v) \right)|_{q=q(Q),p=p(Q,P),v=v(Q,V)} =
{\rm det}\left(\frac{\partial Q_b^{n_b}}{\partial q_a^{n_a}} \right)^{-1}{\rm
\delta}(\Gamma_b), \\
{\rm det}\frac{\partial^2L_{\rm q}}{\partial v_a\partial
v_b}|_{q=q(Q),v=v(Q,V)} ={\rm det}^{2}\left(\frac{\partial Q_a^{n_a}}{\partial
q_b^{n_b}} \right){\rm det}\frac{\partial^2 L_{\rm Q}}{\partial V_a \partial
V_b},
\end{eqnarray}
we get
\begin{eqnarray}
Z_{\rm g} &=& \int {\cal D}Q_a^{r_a}{\cal D}P_a^{r_a}{\cal D}V_a{\rm
\delta}(\Gamma_a){\rm det}\frac{\partial^2L_{\rm Q}}{\partial V_a\partial V_b}
\nonumber \\
& & {} \times \exp i\int dt [P_a^{s_a}(\dot{Q}_a^{s_a} - \bar{Q}_a^{s_a}) +
P_a^{n_a}(\dot{Q}_a^{n_a} - V_a) + L_{\rm Q}].
\end{eqnarray}
This shows that
\begin{equation}
Z_{\rm g}=Z_{\rm O}=Z_{\rm D}=Z_{\rm G}.
\end{equation}
We have found that the generalized canonical formalism is equivalent to the
Ostrogradski's one and these two formalisms are connected by a canonical
transformation.

{\bf Singular case}

First, we show the equivalence between the path integrals $Z_{\rm Ds}$ given by
 (\ref{eq:ZDS}) and $Z_{\rm gs}$ constructed from the Lagrangian $L_{\rm g}$ in
(\ref{eq:LSg}).
In this case, we choose, without loss of generality, for extra constraints the
following form:
\begin{equation}
\omega_A(q^r,p^{n-1},v) \stackrel{\rm def}{\equiv} p_A^{n_A-1} - \frac{\partial
L_{\rm q}}{\partial q_A^{n_A}} + \frac{\partial^2 L_{\rm q}}{\partial
v_A\partial q_b^{n_b}}v_b + \frac{\partial^2 L_{\rm q}}{\partial v_A\partial
q_b^{s_b}}q_b^{s_b+1} \approx 0. \label{eq:sgomega}
\end{equation}
By introducing additional multipliers $\lambda_A^{(5)}$, the Hamiltonian is
given by
\begin{equation}
\bar{H}_{\rm gs}=H_{\rm g}(q^r,p^r) + \lambda_a^{(1)r_a}\pi_a^{r_a} +
\lambda_a^{(2)r_a}\psi_a^{r_a} + \lambda_a^{(3)}\theta_a +
\lambda_a^{(4)}\gamma_a + \lambda_A^{(5)} \omega_A.
\end{equation}
All the constraints $\sigma_\alpha^{\rm (s)} \stackrel{\rm def}{\equiv}
(\theta_a,\psi_a^{r_a},\pi_a^{r_a},\gamma_a,\omega_A)$ give for the determinant
of the matrix $\left(\{\sigma_\alpha^{\rm (s)},\sigma_\beta^{\rm (s)} \}_{\rm
P} \right)$
\begin{equation}
{\rm det}\{\sigma_\alpha^{\rm (s)},\sigma_\beta^{\rm (s)} \}_{\rm P}={\rm det}
\pmatrix{
\displaystyle 0 & \displaystyle -\frac{\partial^2L_{\rm q}}{\partial
v_a\partial v_b} & \displaystyle -\frac{\partial \omega_A}{\partial v_a} \cr
\displaystyle \frac{\partial^2L_{\rm q}}{\partial v_a \partial v_b} &
\displaystyle\{\gamma_a,\gamma_b \}_{\rm P} & \displaystyle\{\gamma_a,\omega_B
\}_{\rm P} \cr
\displaystyle\frac{\partial \omega_A}{\partial v_b} &
\displaystyle\{\omega_A,\gamma_b \}_{\rm P} & \displaystyle\{\omega_A,\omega_B
\}_{\rm P} \cr
}. \label{eq:det}
\end{equation}
If this is nonzero, all the Lagrange multipliers are determined. The path
integral is given by
\begin{eqnarray}
Z_{\rm gs} &=& \int {\cal D}q_a^{r_a}{\cal D}p_a^{r_a}{\cal D}\mu_a^{r_a}{\cal
D}\pi_a^{r_a}{\cal D}v_a{\cal D}\theta_a{\rm \delta}(\pi^r){\rm
\delta}(\psi^r){\rm \delta}(\theta){\rm \delta}(\gamma){\rm \delta}(\omega_A)
{\rm det}^{\frac{1}{2}}\{\sigma_\alpha^{\rm (s)},\sigma_\beta^{\rm (s)} \}_{\rm
P} \nonumber \\
& & {} \times \exp i \int dt [p_a^{r_a}\dot{q}_a^{r_a} +
\pi_a^{r_a}\dot{\mu}_a^{r_a} + \theta_a\dot{v}_a - H_{\rm gs}].
\end{eqnarray}
Integrations with respect to $\mu_a^{r_a},\pi_a^{r_a}$ and $\theta_a$ give
\begin{eqnarray}
Z_{\rm gs} &=& \int {\cal D}q_a^{r_a}{\cal D} p_a^{r_a}{\cal D}v_a{\rm
\delta}\left(\gamma_a(q^r,p^n,v) \right){\rm
\delta}\left(\omega_A(q^r,p^{n-1},v) \right){\rm
det}^{\frac{1}{2}}\{\sigma_\alpha^{\rm (s)},\sigma_\beta^{\rm (s)} \}_{\rm P}
\nonumber \\
& &{} \times \exp i \int dt [p_a^{s_a}(\dot{q}_a^{s_a} -
q_a^{s_a+1})+p_a^{n_a}(\dot{q}_a^{n_a} - v_a) + L_{\rm q}(q,v)]. \label{eq:135}
\end{eqnarray}
Here, we consider the matrix $\left(\{\sigma_\alpha^{\rm (s)},\sigma_\beta^{\rm
(s)} \}_{\rm P} \right)$ . We change this into a form which can be integrated
with respect to $v_a$. The assumption that the determinant of this matrix is
nonzero means
\begin{equation}
{\rm rank}\frac{\partial \omega_B}{\partial v_a}=\rho.
\end{equation}
In the matrix
\begin{equation}
\pmatrix {
\displaystyle\frac{\partial^2L_{\rm q}}{\partial v_a\partial v_b} &
\displaystyle\frac{\partial \omega_A}{\partial v_a} \cr
}=\frac{\partial }{\partial v_a} \pmatrix{\gamma_b & \omega_A \cr },
\end{equation}
we select $\gamma_\xi \; (\xi=\rho+1,\cdots,N)$ which satisfy
\begin{equation}
{\rm det}\left(\frac{\partial (\gamma_\xi,\omega_A)}{\partial v_a} \right)
\not=0,
\end{equation}
to define as $\Xi_a(q^r,p^n,p^{n-1}) \stackrel{\rm def}{\equiv}
(\gamma_\xi,\omega_A)$. The determinant of the matrix (\ref{eq:det}) reduces to
\begin{equation}
{\rm det}\{\sigma_{\alpha}^{\rm (s)},\sigma_\beta^{\rm (s)} \}_{\rm P}={\rm
det}^2\left(\frac{\partial \Xi_a}{\partial v_a} \right){\rm
det}\{\gamma_A,\gamma_B \}_{\rm P}.
\end{equation}
Then the path integral (\ref{eq:135}) is given by
\begin{eqnarray}
Z_{\rm gs} &=& \int{\cal D}q_a^{r_a}{\cal D}p_a^{r_a}{\cal D}v_a{\rm
\delta}(\gamma_A){\rm \delta}(\Xi_a){\rm det}\left(\frac{\partial
\Xi_a}{\partial v_b} \right){\rm det}^{\frac{1}{2}}\{\gamma_A,\gamma_B \}_{\rm
P} \nonumber \\
& &{} \times\exp i \int dt [p_a^{s_a}(\dot{q}_a^{s_a} - q_a^{s_a+1}) +
p_a^{n_a}(\dot{q}_a^{n_a} - v_a) +L_{\rm q}].
\end{eqnarray}
Integrations with respect to $v_a,p_a^{s_a}$ and $q_a^{s_a+1}$ give
\begin{equation}
Z_{\rm gs}=\int{\cal D}q_a^1{\cal D}p_a^{n_a}{\rm \delta}\left(\gamma_A
\right){\rm det}^{\frac{1}{2}}\{\gamma_A,\gamma_B \}_{\rm P} \exp i \int dt
[p_a^{n_a}q_a^{1(n_a)} - \hat{H}_{\rm gs}], \label{eq:Zgs1}
\end{equation}
where
\begin{equation}
\hat{H}_{\rm gs} \stackrel{\rm def}{\equiv} p_a^{n_a}v_a - L_{\rm
q}\left(q^1,q^{1(s)},v \right).
\end{equation}
Putting $\gamma_A=\phi_A$, we have arrived at the same expression as $Z_{\rm
Ds}$ in (\ref{eq:ZDS}).

Next task is canonical transformation. Since the exponent in (\ref{eq:ZGS}) is
the same as in Eq. (\ref{eq:ZG}), we insert Eq. (\ref{eq:unity}) into the
expression (\ref{eq:135}) and integrate with respect to $q^r,p^r$ and $v$ to
obtain
\begin{eqnarray}
Z_{\rm gs} &=& \int {\cal D}Q^r{\cal D}P^r{\cal D}V{\rm \delta}(\gamma_a){\rm
\delta}(\omega_A){\rm det}^{\frac{1}{2}}\{\sigma_\alpha^{\rm
(s)},\sigma_\beta^{\rm (s)} \}_{\rm P} \nonumber \\
& &{}\times {\rm det}\left(\frac{\partial Q_b^{n_b}}{\partial
q_a^{n_a}}\right)^{-1}|_{q=q(Q),p=p(Q,P),v=v(Q,V)} \nonumber \\
& &{}\times \exp i \int dt [p_a^{s_a}(\dot{Q}_a^{s_a} - \bar{Q}_a^{s_a}) +
P_a^{n_a}(\dot{Q}_a^{n_a} - V_a) + L_{\rm Q}].
\end{eqnarray}
By using the relations
\begin{eqnarray}
{\rm \delta}(\gamma_a)|_{q=q(Q),p=p(Q,P),v=v(Q,V)} ={\rm \delta}(\Gamma_a){\rm
det}\left(\frac{\partial Q_b^{n_b}}{\partial q_a^{n_a}} \right)^{-1}|_{q=q(Q)},
\\
{\rm \delta}(\omega_A)|_{q=q(Q),p=p(Q,P),v=v(Q,V)}={\rm \delta}(\Omega_A){\rm
det}\left(\frac{\partial Q_B^{n_B}}{\partial q_A^{n_A}} \right)^{-1}|_{q=q(Q)},
\\
{\rm det}\{\sigma_\alpha^{\rm (s)},\sigma_\beta^{\rm (s)} \}_{\rm P}={\rm
det}^4\left(\frac{\partial Q_b^{n_b}}{\partial q_a^{n_a}} \right){\rm
det}^2\left(\frac{\partial Q_B^{n_B}}{\partial q_A^{n_A}} \right){\rm
det}\{\Sigma_\alpha^{\rm (s)},\Sigma_{\beta}^{\rm (s)} \}_{\rm P},
\end{eqnarray}
we obtain
\begin{eqnarray}
Z_{\rm gs} &=& \int{\cal D}Q^r{\cal D}P^r{\cal D}V[{\rm \delta}(\Gamma_a){\rm
\delta}(\Omega_A){\rm det}^{\frac{1}{2}}\{\Sigma_\alpha^{\rm
(s)},\Sigma_\beta^{\rm (s)} \}_{\rm P}] \nonumber \\
& &{} \times \exp i \int dt [P_a^{s_a}(\dot{Q}_a^{s_a} - \bar{Q}_a^{sa}) +
P_a^{n_a}(\dot{Q}_a^{n_a} - V_a) + L_{\rm Q}].
\end{eqnarray}
This shows
\begin{equation}
Z_{\rm gs}=Z_{\rm Os} = Z_{\rm Ds} = Z_{\rm Gs}.
\end{equation}
The path integrals $Z_{\rm gs}$ and $Z_{\rm Gs}$ are connected with each other
by the canonical transformation generated by $F$ in (\ref{eq:F}).

\section{Summary and Discussion}

In the present paper we have given path integral expressions for three
canonical formalisms of higher-derivative theories. For each formalism we have
considerd both nonsingular and singular cases. It has been shown that three
formalisms share the same path integral expressions. In paticular it has been
pointed out that the generalized canonical formalism is canonically
transformated from the constrained canonical one.

Here we have to mention some crucial properties involved in higher-derivative
theories. The Hamiltonian is unbounded from below in general; unitarity is
violated in general; whether or not stable vacuum can be well defined is
problematic.
That means we should worry about how to define path integral.
Leaving these problems to the future investigation, we have just assumed in
this paper that stable lowest state can be defined, and the path integral can
be written down as usual by the use of a time development operator, the
Hamiltonian.

\section*{Acknowledgements}

The authors would like to thank C. Dariescu, M.-A. Dariescu, M. Hirayama, T.
Kurimoto, K. Matumoto and H.Yamakoshi for discussions.

\section*{References}
1) D. A. Eliezer and R. P. Woodard, Nucl. Phys. {\bf B325} (1989) 389.

2) J. Z. Simon, Phys. Rev. {\bf D41} (1990) 3720.

3) K. S. Stelle, Phys. Rev. {\bf D16} (1977) 953.

4) M. Ostrogradski, Mem. Ac. St. Petersbourg {\bf V14} (1850) 385.

5) P. A. M. Dirac, {\it Lectures on Quantum Mechanics} (Yeshiva University
Press, New York, 1964).

6) J. M. Pons, Lett. Math. Phys. {\bf 17} (1989) 181.

7) C. Batlle, J. Gomils, J. M. Pons and N. Roman-Roy, J. Phys. {\bf  A21}
(1988) 2693.

8) J. Govaerts and M. S. Rashid, hep-th/9403009.

9) I. L. Buchbinder and S. L. Lyahovich, Class. Quantum Grav. {\bf 4} (1987)
1487.

\end{document}